\documentclass[usenatbib]{mn2e}
\usepackage{epsfig}

\newcommand{\etal}{et~al.}  
\newcommand{\ionhy}{H{\sc ii}}
\newcommand{\glim}{{\em GLIMPSE}}
 
\newcommand{\kms}{$\mbox{km~s}^{-1}$}
\newcommand{\transa}{$5_{1}\mbox{-}6_{0}\mbox{~A}^{+}$}
\newcommand{\water}{$\mbox{H}_{2}\mbox{O}$}

\newcommand{\specdfig}[2]        
{
  \begin{center}
    \begin{minipage}[t]{0.45\textwidth}
        \psfig{file=#1.eps,height=\textwidth,angle=270}
    \end{minipage}
    \hfill
    \begin{minipage}[t]{0.45\textwidth}
        \psfig{file=#2.eps,height=\textwidth,angle=270}
    \end{minipage}
  \end{center}
}

\newcommand{\specsfig}[1]        
{
  \begin{center}
    \begin{minipage}[t]{0.45\textwidth}
        \psfig{file=#1.eps,height=\textwidth,angle=270}
    \end{minipage}
  \end{center}
}

\begin{document}

\title[A \glim-based search for Methanol Masers] 
{A \glim-based search for 6.7-GHz Methanol Masers and the lifetime of their spectral features}

\author[Ellingsen]{S.P. Ellingsen$^1$\\
$^1$ School of Mathematics and Physics, University of Tasmania, 
     Private Bag 21, Hobart, Tasmania 7001, Australia;\\  
     Simon.Ellingsen@utas.edu.au}

\maketitle

\begin{abstract}

  The University of Tasmania Mt Pleasant 26-m and Ceduna 30-m
  radio telescopes have been used to search for 6.7-GHz class~II
  methanol masers towards two hundred \glim\ sources.  The target
  regions were selected on the basis of their mid-infrared colours as
  being likely to be young high-mass star formation regions and are
  either bright at 8.0~\micron\/, or have extreme [3.6]-[4.5] colour.
  Methanol masers were detected towards 38 sites, nine of these being
  new detections.  The prediction was that approximately 20 new
  6.7-GHz methanol masers would be detected within 3.5~arcmin of the
  target {\em GLIMPSE} sources, but this is the case for only six of
  the new detections.  A number of possible reasons for the
  discrepancy between the predicted and actual number of new
  detections have been investigated.  It was not possible to draw any
  firm conclusions as to the cause, but it may be because many of the
  target sources are at an evolutionary phase prior to that associated
  with 6.7-GHz methanol masers.  Through comparison of the spectra
  collected as part of this search with those in the literature, the
  average lifetime of individual 6.7-GHz methanol maser spectral
  features is estimated to be around 150 years, much longer than is
  observed for 22-GHz water masers.
 
\end{abstract}


\section{Introduction} \label{sec:intro}

In the forty years since their discovery, molecular masers have proven
to be powerful signposts of star formation within our Galaxy.  The
6.7-GHz \transa\/ transition of methanol is the second strongest of
all molecular masers (after the 22-GHz \water\/ transition) and has
been shown to trace high-mass star formation regions from a very early
evolutionary phase until soon after the formation of an ultra-compact
\ionhy\/ region \citep[e.g.][]{E06,MBH+05}.  It also has the advantage
that unlike OH and \water\/ masers, 6.7-GHz methanol masers are only
associated with high-mass star formation \citep{MENB03}.  To date
approximately 520 6.7-GHz methanol masers have been detected
\citep*{PMB05}, using three basic strategies; searches targeted
towards main-line OH and 12.2-GHz methanol masers
\citep*[e.g.][]{M91,MGN92,MG92,CVEWN95}, searches targeted towards {\em IRAS}
sources with colours characteristic of ultra-compact \ionhy\/ regions
\citep*[e.g.][]{SVGM93,VGM95,WHRB97,SVK+99,SK00} and blind searches of
some regions of the Galactic Plane \citep[e.g.][]{C96,EVM+96,SKHKP02}.

Searches towards sources selected on the basis of their {\em
IRAS}-colours detected many new 6.7-GHz methanol masers, however, they
also suffered from a number of shortcomings, in particular false
associations and incompleteness.  Subsequent observations with better
positional accuracy show that in some cases the masers are not
associated with the {\em IRAS} source towards which the search was
made (false association).  From the point of view of finding new
methanol maser sites this is irrelevant, the {\em IRAS} point source
catalogue has led to targeted observations of a high-mass star
formation region containing masers.  However, it does mean that it is
pointless trying to infer much about the physical conditions in the
masing region from {\em IRAS} measurements.  This is true even for
sources where the masers are associated with the {\em IRAS} source
because the large beamwidth of the {\em IRAS} observations means that
the infrared flux densities measured often include contributions from
multiple star formation sites within the larger molecular cloud.  The
second problem with {\em IRAS}-based searches is that they fail to
detect many maser sites.  The untargeted search of \citet{EVM+96}
approximately doubled the number of methanol masers in the regions
searched, despite the fact that a number of searches targeted towards
OH masers or {\em IRAS} sources had already been undertaken.
\citet{EVM+96} showed that no {\em IRAS}-based search was able to
detect more than 60\% of the methanol masers in a given region due to
many of the maser sites having no associated {\em IRAS} source.  The
absence of an {\em IRAS} source towards these methanol maser sites is
likely due to well-documented problems with confusion and hyteresis in
{\em IRAS} observations of the Galactic Plane and not because there
is no far-infrared emission associated with the masers.

In the 20 years since the {\em IRAS} observations a number of other
infrared satellites have made large imaging surveys of the Galactic
Plane, in particular {\em MSX} \citep{E+03} and the {\em GLIMPSE}
survey with the {\em Spitzer Space Telescope} \citep{BCB+03}.  The
{\em MSX} observations covered 4 bands from 8 - 21~\micron\/ with a
resolution of 18~arcsec at 21~\micron.  The {\em GLIMPSE} survey was
at still shorter wavelengths (4 bands from 3.6 through to
8.0~\micron), but at much higher spatial resolution (1.4--1.9~arcsec).
The {\em MSX} observations are not sufficiently sensitive, nor of high
enough angular resolution to provide useful targeting criteria for
maser searches \citep{E05}.  Previous ground-based mid-infrared
observations have found that many methanol maser sources had no
associated source at 10~\micron\/ \citep{DPT00,WBBN01} and so the
prospects for the {\em GLIMPSE} survey being a useful tool for maser
studies might seem poor.  However, the higher sensitivity of the
space-based observations means that this is not the case and
\citet{E06} (hereafter E06) found that more than 90\% of 6.7~GHz
methanol maser sites are associated with mid-infrared emission visible
in the 8.0~\micron\/ band.  E06 showed that approximately two-thirds
of methanol maser sites are associated with a {\em GLIMPSE} point
source and that typically these sources are bright in the
8.0~\micron\/ band and have very red mid-infrared colours.

Analysing previous untargeted searches of the Galactic Plane, E06
found that approximately 10\% of {\em GLIMPSE} point sources which
have 8.0-\micron\/ intensity less than tenth magnitude ([8.0] $<$ 10)
and colour between the 3.6- and 4.5-\micron\/ bands greater than 1.3
magnitudes ([3.6]-[4.5] $>$ 1.3) were associated with methanol masers.
They also found that such a search of all sources meeting these {\em
  GLIMPSE}-based criteria within a given region would detect more than
80\% of all the methanol masers in the region (some of these would be
serendipitous detections where the maser is not associated with a
targeted {\em GLIMPSE} source, but merely nearby).  The {\em GLIMPSE}
point source catalogue contains more than 30 million sources and of
these 5675 meet the criteria outlined by E06.  Statistically, these
cannot all be young high-mass star forming regions, and there must be
contamination of the sample by other objects (E06 suggested low-mass
class~0 objects as one example).  Clearly it is desirable to
determine if it is possible to further refine the sample to increase
the percentage of targeted sources that have an associated 6.7-GHz
methanol maser.  Many of the {\em GLIMPSE} point sources that meet the
criteria lie very close to one or another of the cut-offs and it may
well be that the percentage of contaminating sources increases 
closer to the criterion limits.  To test this hypothesis I selected
the 100 most extreme sources meeting each criterion and searched for
6.7-GHz methanol maser emission towards these sources.

Despite excluding {\em GLIMPSE} sources close to known 6.7-GHz
methanol masers from the target sample (see section~\ref{sec:obs}),
the redetection of some previously detected masers is inevitable in
any large search.  The majority of the currently known 6.7-GHz
methanol masers were discovered more than a decade ago and so
comparison of a current epoch spectrum with those in the literature
provides an opportunity to assess the long-term variability of 6.7-GHz
methanol masers.  The variability of a large sample of 6.7-GHz
methanol masers on timescales of a few months to a year was
investigated by \citet{CVE95}.  They found the variability on a
timescale of a year is typically less than a factor of two, but
sometimes is more extreme.  They also found greater variability was
more prevalent in masers with weaker peak flux density.  \citet{GGV04}
observed a sample of 54 sources regularly for more than four years and
observed a range of different types of variability, including periodic
variations which are not observed in water or OH masers in star
formation regions.  These observations showed significant variability
in more than half of the spectral features analysed, suggesting the
sparse sampling of \citet{CVE95} underestimated the true variability.

22-GHz water masers are known to exhibit much more extreme variability
than is observed in 6.7-GHz methanol masers, with many individual
spectral features having lifetimes of around one year \citep{BCC+03}.
\citet{GGV04} found only one of 54 6.7-GHz methanol masers (Mon~R2)
showed significant changes in its spectral profile, similar to the
variations commonly observed in water masers.  \citet{V05} estimates
the lifetime of 6.7-GHz methanol maser sources to be in the range
2.5--4.5 $\times 10^4$ years.  Individual spectral features might be
expected to have significantly shorter lifetimes, and by comparing how
many spectral features have disappeared and appeared over a ten year
period for a sample of sources it should be possible to estimate the
average lifetime of typical 6.7-GHz methanol maser spectral
features/spots.

\section{Observations and Data reduction} \label{sec:obs}

\begin{figure}
  \psfig{file=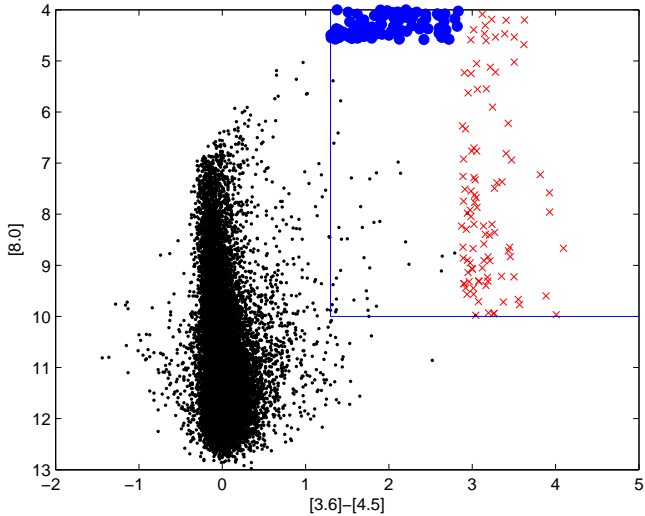,height=0.425\textwidth}
  \caption{A [3.6]-[4.5] versus [8.0] colour-magnitude diagram for the
    {\em GLIMPSE} point sources searched for 6.7-GHz methanol masers.
    The crosses are the sources with red [3.6]-[4.5] colour (category
    A), the circles are the sources with intense 8.0-\micron\/
    emission (category B) and the dots are the {\em GLIMPSE} point
    sources within 30~arcsec of $\ell$=326.5\degr, $b$ = 0.0\degr\/
    which have measurements in all of the 3.6-, 4.5- and 8.0-\micron\/
    bands (21992 of 109053 sources).  The box marks the borders of the
    region identified by E06 as containing sources likely to be young
    high-mass star forming regions.}
  \label{fig:selection}
\end{figure}

To test the predictions of E06 I have carried out a search for 6.7-GHz
methanol masers towards 200 {\em GLIMPSE} point sources.  To do this I
started with the sample of 5675 {\em GLIMPSE} point sources that meet
both criteria [8.0] $<$ 10 and [3.6]-[4.5] $>$ 1.3.  The sample was
drawn from the 2005 April 15 release of the {\em GLIMPSE} point source
catalogue and point sources without measurements in all three of these
bands were excluded. From this sample of 5675 sources I removed those
that lay within 3.5~arcmin of a methanol maser source listed in the
\citet{PMB05} catalogue (3.5~arcmin is half the FWHM of the Mt
Pleasant radio telescope at 6.7~GHz).  This left a total of 4878 {\em
  GLIMPSE} point sources and from these I extracted the 100 sources
strongest in the 8.0-\micron\/ band and the 100 sources with the
largest [3.6]-[4.5] colour.  The hundred sources selected on the basis
of their 8.0-\micron\/ strength have an intensity of 4.58 magnitudes
or less (hereafter referred to as category B sources) and the 100
reddest sources have [3.6]-[4.5] colour greater than 2.87 magnitudes
(category A).  Figure~\ref{fig:selection} shows a colour-magnitude
diagram of the selected sources, compared to all sources within a
30-arcsecond radius of $\ell$ = 326.5\degr, $b$ = 0.0\degr.  This
shows that the selected sources have very different properties to the
majority of {\em GLIMPSE} sources.

The observations were made between 2005 July-December using the
University of Tasmania Mt Pleasant 26m and Ceduna 30m radio
telescopes.  A rest frequency of 6.668518~GHz for the \transa\/
transition was used in the observations and data processing.  A more
precise determination of the rest frequency determine by \cite{BK95}
measured it to be 6.6685192(8)~GHz, and the frequency difference of
1.2~kHz corresponds to a velocity change of 0.05~\kms.  As this is
less than the accuracy to which the peak velocity can easily be
determined, no correction has been made to the velocity scale to
account for this.  At 6.7~GHz the half-power beam width of the Mt
Pleasant and Ceduna antennas are 7 and 6~arcmin respectively.  The
majority of the initial search observations were made at Mt Pleasant,
as were all the final spectra (except G345.20-0.04).  Ceduna was used
for some repeat observations (where the initial observations were
affected by internally generated interference) and to determine better
positions for some sources.  The Mt Pleasant observations were made
using a cryogenically cooled receiver with dual orthogonal circular
polarizations each with a system temperature which varied between
750-1400~Jy depending on the day.  The data were collected using a
2-bit digital autocorrelation spectrometer configured with 4096
channels spanning a 4-MHz bandwidth.  For an observing frequency of
6.7~GHz this configuration yields a natural weighting (i.e. no Hanning
or other smoothing) velocity resolution of 0.044~\kms.  The Ceduna
observations were made with an uncooled receiver with dual orthogonal
circular polarizations each with a system temperature of approximately
500~Jy.  The Ceduna correlator was identical to that used at Hobart,
however due to technical problems only 2048 channels were available
meaning that the natural weighting velocity resolution was 0.105~\kms.

Each \glim\ source was observed for 10 minutes on-source, using an
observation of a nearby target for the off-source reference.  This
yielded a typical RMS between 0.5 and 0.6~Jy after the spectrum had
been smoothed with 5-point box car (giving an effective velocity
resolution of 0.22~\kms).  Some class~II methanol masers exhibit very
narrow spectral features with \citet{SK00} showing the median velocity
full-width half maximum of the methanol masers they observed to be
0.14~\kms\/.  To avoid missing weaker masers with very narrow features
spectra were also produced for each source with only Hanning smoothing
applied.  This yields a velocity resolution of 0.088~\kms\/ and a
typical RMS between 0.8 and 0.9 Jy.  For each of the spectra showing
likely maser emission, a 5-point grid observation (with observations
made at 4 points forming a square, each offset from the fifth, central
position by 3.5~arcminutes) was made to better determine the maser
position.  A non-linear least squares fit of a 2-dimensional Gaussian
with the same dimensions as the telescope beam was made to the data
from the grid observations in order to better determine the location
of the maser emission.  In a few cases the masers were significantly
offset from the targeted {\em GLIMPSE} source and the process had to
be repeated to refine the position.  Once an accurate position had
been determined a final, sensitive observation of all the detected
sources was then made.  A 5-point grid observation was also made at
the same time as the final spectrum and the measured positional offset
used to correct the flux density scale for primary beam attenuation
(typically this correction was less than 10 per cent).  Through
comparison of the measured and known positions for sources that have
been previously detected the pointing accuracy of the system is
estimated to be approximately 0.5~arcmin.

\section{Results} \label{sec:results}

A search of 200 {\em GLIMPSE} sources has resulted in the detection of
thirty eight 6.7-GHz methanol masers, nine of these being new
discoveries (see Table~\ref{tab:det}).  Methanol maser emission was
detected in the spectra of 27 {\em GLIMPSE} sources, with nine of
these containing two separate sites and one three, leading to a total
of 38 detections.  Table~\ref{tab:det} lists the 6.7-GHz methanol
maser sources detected, their location and the {\em GLIMPSE} source
they were detected towards.  Sources with no entry in the {\em
  GLIMPSE} column were typically detected when making grid or final
observations of the previous source in the table.
Table~\ref{tab:maserprop} summarises the characteristics of the maser
emission, including the peak flux density, velocity range and
integrated flux density of the sources.

\begin{table*}
  \caption{6.7-GHz methanol maser sources detected towards {\em GLIMPSE}
    sources.  Sources marked with a $^*$ are new detections.  See the
    catalogue of \citet{PMB05} for the references for previously
    detected sources.  Those listed as being in category
    A were selected on the basis of having a large [3.6]-[4.5] excess,
    while those in category B were selected on the basis of high
    intensity in the 8.0~\micron\/ band. } 
  \begin{tabular}{lllllr} \hline
    {\bf Methanol} & {\bf Right}     & {\bf Declination} & {\bf \glim}    &
      {\bf Category} & {\bf Angular}    \\ 
    {\bf maser}    & {\bf Ascension} &                   & {\bf PSC name} &
                     & {\bf Separation} \\
    {\bf name}     & {\bf (J2000)}   & {\bf (J2000)}     &                &
                     & {\bf (arcmin)}   \\ [2mm] \hline
    G\,$305.21\!+\!0.21$   & 13:11:14.4 & -62:34:26 & GLMC G\,$305.4450\!+\!0.2644$ & A &  14.5 \\
    G\,$305.20\!+\!0.21$   & 13:11:10.6 & -62:34:39 &                               &   &       \\
    G\,$316.81\!-\!0.06$   & 14:45:26.4 & -59:49:17 & GLMC G\,$316.7626\!-\!0.0128$ & A &   4.0 \\
    G\,$317.47\!-\!0.41^*$ & 14:51:22.5 & -59:51:07 & GLMC G\,$317.4656\!-\!0.4026$ & A &   0.5 \\ 
    G\,$320.23\!-\!0.29$   & 15:09:52.0 & -58:25:38 & GLMC G\,$320.3686\!-\!0.2804$ & A &   8.2 \\
    G\,$323.46\!-\!0.08$   & 15:29:19.3 & -56:31:23 & GLMC G\,$323.4702\!-\!0.1421$ & B &   3.9 \\
    G\,$323.74\!-\!0.26$   & 15:31:45.4 & -56:30:50 & GLMC G\,$323.5478\!+\!0.0237$ & A &  20.7 \\
    G\,$326.66\!+\!0.52$   & 15:45:02.9 & -54:09:03 & GLMC G\,$326.7077\!+\!0.5828$ & A &   4.6 \\ 
    G\,$326.64\!+\!0.61$   & 15:44:33.2 & -54:05:31 &                               &   &       \\
    G\,$328.24\!-\!0.55$   & 15:57:58.3 & -53:59:23 & GLMC G\,$328.2558\!-\!0.4124$ & A &   8.2 \\
    G\,$328.25\!-\!0.53$   & 15:57:59.9 & -53:58:01 &                               &   &       \\
    G\,$331.56\!-\!0.12$   & 16:12:27.2 & -51:27:37 & GLMC G\,$331.6047\!-\!0.0724$ & B &   4.1 \\
    G\,$331.54\!-\!0.07$   & 16:12:08.8 & -51:25:47 &                               &   &       \\
    G\,$337.15\!-\!0.04^*$ & 16:37:50.9 & -47:39:09 & GLMC G\,$337.1393\!-\!0.3803$ & A &   1.6 \\ 
    G\,$337.39\!-\!0.20^*$ & 16:37:55.9 & -47:20:58 & GLMC G\,$337.3444\!-\!0.1807$ & A &   3.2 \\ 
    G\,$339.88\!-\!1.26$   & 16:52:04.7 & -46:08:34 & GLMC G\,$339.8712\!-\!0.6697$ & B &  35.4 \\
    G\,$340.97\!-\!1.03^*$ & 16:54:58.7 & -45:09:23 & GLMC G\,$340.9108\!-\!1.0316$ & A &   3.5 \\
    G\,$343.52\!-\!0.50^*$ & 17:01:27.7 & -42:49:44 & GLMC G\,$343.5015\!-\!0.4724$ & A &   1.7 \\
    G\,$345.20\!-\!0.04^*$ & 17:05:01.0 & -41:13:09 & GLMC G\,$345.2563\!-\!0.0366$ & A &   3.6 \\
    G\,$11.15\!-\!0.14^*$  & 18:10:28.1 & -19:22:40 & GLMC G\,$011.2013\!-\!0.1570$ & A &   6.2 \\
    G\,$18.99\!-\!0.04^*$  & 18:25:44.1 & -12:24:15 & GLMC G\,$019.0087\!-\!0.0293$ & A &   1.5 \\
    G\,$23.01\!-\!0.41$    & 18:34:40.2 & -09:00:36 & GLMC G\,$022.9600\!-\!0.4752$ & B &   4.9 \\
    G\,$24.14\!+\!0.00$    & 18:35:20.8 & -07:48:48 & GLMC G\,$024.2356\!-\!0.0570$ & B &   5.6 \\
    G\,$25.82\!-\!0.17$    & 18:39:03.6 & -06:24:10 & GLMC G\,$025.8062\!-\!0.2845$ & B &   6.5 \\
    G\,$27.28\!+\!0.15$    & 18:40:34.5 & -04:57:14 & GLMC G\,$027.2788\!+\!0.2095$ & B &   3.6 \\
    G\,$27.21\!+\!0.26$    & 18:40:03.8 & -04:58:09 &                               &   &      	\\  
    G\,$28.02\!-\!0.44$    & 18:44:02.1 & -04:34:14 & GLMC G\,$027.9676\!-\!0.4747$ & A &   3.8 \\
    G\,$29.86\!-\!0.04$    & 18:45:59.5 & -02:44:47 & GLMC G\,$029.8400\!-\!0.0973$ & B &   3.7 \\
    G\,$29.95\!-\!0.02$    & 18:46:03.7 & -02:39:21 &                               &   &       \\
    G\,$30.20\!-\!0.17$    & 18:47:03.5 & -02:30:31 &                               &   &       \\ 
    G\,$30.70\!-\!0.07$    & 18:47:36.9 & -02:01:05 & GLMC G\,$030.6721\!+\!0.0141$ & A &   5.3 \\
    G\,$30.76\!-\!0.05$    & 18:47:39.7 & -01:57:22 &                               &   &       \\
    G\,$30.79\!+\!0.20$    & 18:46:48.1 & -01:48:46 & GLMC G\,$030.8070\!+\!0.0803$ & B &   7.5 \\
    G\,$30.78\!+\!0.23$    & 18:46:41.5 & -01:48:32 &                               &   &       \\
    G\,$30.82\!-\!0.05$    & 18:46:37.4 & -01:45:14 &                               &   &       \\
    G\,$31.04\!+\!0.36$    & 18:46:41.5 & -01:30:42 & GLMC G\,$031.0431\!+\!0.2820$ & A &   5.1 \\ 
    G\,$35.20\!-\!0.74$    & 18:58:12.7 & +01:40:50 & GLMC G\,$035.1293\!-\!0.7427$ & A &   4.2 \\ 
    G\,$35.18\!-\!0.74^*$    & 18:58:09.8 & +01:39:36 &                               &   &       \\ \hline
  \end{tabular}
  \label{tab:det}
\end{table*}    

An oversight during the formulation of the sample meant that there
were two cases where two {\em GLIMPSE} sources with only a small
separation between them were observed (these being
GLMC~G\,$321.9359\!-\!0.0061$ \& GLMC~G\,$321.9367\!-\!0.0063$ and
GLMC~G\,$345.2563\!-\!0.0366$ \& GLMC~G\,$345.2574\!-\!0.0351$ ).  So
effectively a total of 198 locations were searched for 6.7-GHz
methanol masers in this work.  Table~\ref{tab:nondet} lists the {\em
GLIMPSE} point sources observed which contained no detected 6.7-GHz
methanol maser emission.  For each source it gives the velocity range
covered by the spectrum and the RMS in the smoothed (velocity
resolution 0.22~\kms) spectrum.

\begin{table*}
  \caption{Properties of the 6.7-GHz methanol maser sources detected.} 
  \begin{tabular}{lrrcr} \hline
    {\bf Methanol} & {\bf Peak Flux}    & {\bf Peak}   & {\bf Velocity} &
      {\bf Integrated}   \\ 
    {\bf maser}    & {\bf Density}      & {\bf Vel.}   & {\bf range}    &
       {\bf Flux Density} \\
    {\bf name}     & {\bf (Jy)}         & {\bf (\kms)} & {\bf (\kms)}   &
      {\bf (Jy\kms)} \\ [2mm] \hline
    G\,$305.21\!+\!0.21$ & 488   & -38.5 & -39.8 -- -34.6  &  377.8 \\ 
    G\,$316.81\!-\!0.06$ & 71    & -45.8 & -47.2 -- -42.0  &   74.1 \\ 
    G\,$317.47\!-\!0.41$ & 61    & -37.7 & -42.2 -- -36.9  &  109.1 \\ 
    G\,$320.23\!-\!0.29$ & 30    & -62.2 & -71.1 -- -59.0  &   45.8 \\ 
    G\,$323.46\!-\!0.08$ & 18    & -67.0 & -67.3 -- -66.6  &    9.7 \\ 
    G\,$323.74\!-\!0.26$ & 3450  & -50.6 & -57.2 -- -42.8  & 5556.9 \\ 
    G\,$326.66\!+\!0.52$ & 15    & -37.3 & -41.4 -- -36.2  &   10.5 \\ 
    G\,$328.24\!-\!0.55$ & 955   & -44.8 & -46.3 -- -33.6  &  560.2 \\ 
    G\,$331.56\!-\!0.12$ & 61    & -97.2 & -104.9 -- -94.3 &   73.2 \\ 
    G\,$337.15\!-\!0.40$ & 24    & -49.4 & -49.8 -- -48.2  &   19.8 \\ 
    G\,$337.39\!-\!0.20$ & 36    & -56.4 & -65.8 -- -52.1  &   51.3 \\ 
    G\,$339.88\!-\!1.26$ & 1629  & -38.8 & -39.9 -- -31.8  & 2898.6 \\ 
    G\,$340.97\!-\!1.03$ & 10    & -31.4 & -32.2 -- -20.1  &    3.7 \\ 
    G\,$343.52\!-\!0.50$ & 4     & -39.0 & -42.3 -- -38.0  &    2.9 \\ 
    G\,$345.20\!-\!0.04$ & 3     & -0.7  & -1.0  -- 0.0    &    2.1 \\ 
    G\,$11.15\!-\!0.14$  & 26    & 32.3  & 18.6  -- 33.7   &   33.4 \\ 
    G\,$18.99\!-\!0.04$  & 20    & 55.4  & 54.5  -- 60.1   &   20.0 \\ 
    G\,$23.01\!-\!0.41$  & 442   & 74.8  & 72.3  -- 83.2   &  806.3 \\ 
    G\,$24.14\!+\!0.00$  & 30    & 17.4  & 17.1  -- 18.0   &   20.6 \\ 
    G\,$25.82\!-\!0.17$  & 78    & 91.6  & 90.0  -- 99.4   &   89.2 \\ 
    G\,$27.28\!+\!0.15$  & 36    & 34.8  & 34.3  -- 35.5   &   18.9 \\ 
    G\,$27.21\!+\!0.26$  & 7     &  9.1  &  7.6  -- 9.8    &    5.5 \\ 
    G\,$28.02\!-\!0.44$  & 4     & 16.8  & 16.0  -- 28.0   &    3.5 \\ 
    G\,$29.86\!-\!0.04$  & 81    & 101.4 & 99.9  -- 104.3  &  112.8 \\ 
    G\,$30.20\!-\!0.17$  & 22    & 108.5 & 104.5 -- 110.9  &   26.4 \\ 
    G\,$30.70\!-\!0.07$  & 154   & 88.2  & 85.7  -- 90.0   &   85.8 \\ 
    G\,$30.79\!+\!0.20$  & 19    & 87.4  & 76.1  -- 102.0  &   51.2 \\ 
    G\,$30.82\!-\!0.05$  & 13    & 91.7  & 90.7  -- 109.3  &   28.3 \\ 
    G\,$31.04\!+\!0.36$  & 4     & 80.7  & 79.2  -- 81.4   &    3.3 \\ 
    G\,$35.18\!-\!0.74$  & 44    & 36.7  & 29.5  -- 37.8   &   49.5 \\ 
    G\,$35.20\!-\!0.74$  & 148   & 28.5  & 27.0  -- 35.0   &  136.8 \\ 
  \end{tabular}
  \label{tab:maserprop}
\end{table*}    

\begin{table*}
  \caption{\glim\ sources searched for 6.7-GHz methanol maser emission
  for which there was no detection.  Those listed as being in category
  A were selected on the basis of having a large [3.6]-[4.5] excess,
  while those in category B were selected on the basis of high
  intensity in the 8.0~\micron\/ band. The RMS quoted is for the
  spectrum after box car smoothing to a velocity resolution of
  0.22~\kms.}
  \begin{tabular}{lcllcc} \hline
    {\bf \glim}    & {\bf Category} & {\bf Right Ascension} & {\bf Declination}
       & {\bf RMS}  & {\bf Velocity Range} \\ 
    {\bf PSC name} &                & {\bf (J2000)}         & {\bf (J2000)}    
       & {\bf (Jy)} & {\bf (\kms)}         \\ [2mm] \hline
    GLMC G\,$301.1149\!+\!0.0094$ & A &  12:35:31.4 & -62:48:22 & 0.5 & -134 -- 45 \\
    GLMC G\,$301.3438\!+\!0.0134$ & A &  12:37:31.5 & -62:48:55 & 0.5 & -134 -- 45 \\
    GLMC G\,$302.4547\!-\!0.7406$ & B &  12:47:08.6 & -63:36:30 & 0.5 & -133 -- 46 \\
    GLMC G\,$302.6482\!+\!0.9860$ & B &  12:49:01.7 & -61:53:04 & 0.5 & -134 -- 45 \\
    GLMC G\,$304.0642\!+\!0.4367$ & A &  13:01:13.2 & -62:24:49 & 0.5 & -133 -- 46 \\
    GLMC G\,$304.6740\!+\!0.2569$ & A &  13:06:34.1 & -62:33:49 & 0.5 & -133 -- 46 \\
    GLMC G\,$305.1384\!-\!0.3755$ & B &  13:11:00.3 & -63:09:48 & 0.5 & -132 -- 47 \\
    GLMC G\,$306.0949\!-\!0.4782$ & B &  13:19:31.8 & -63:10:38 & 0.5 & -132 -- 47 \\
    GLMC G\,$307.8055\!-\!0.4560$ & A &  13:34:27.5 & -62:55:12 & 0.5 & -131 -- 48 \\
    GLMC G\,$308.1289\!-\!0.2989$ & A &  13:37:00.8 & -62:42:35 & 0.5 & -131 -- 48 \\
    GLMC G\,$309.0355\!-\!0.2859$ & B &  13:44:43.9 & -62:31:28 & 0.6 & -131 -- 48 \\
    GLMC G\,$309.1685\!-\!0.5511$ & A &  13:46:20.7 & -62:45:22 & 0.6 & -131 -- 48 \\
    GLMC G\,$310.1545\!+\!0.3379$ & B &  13:52:53.2 & -61:40:09 & 0.6 & -131 -- 48 \\
    GLMC G\,$310.3102\!+\!0.6306$ & A &  13:53:34.3 & -61:20:52 & 0.6 & -131 -- 48 \\
    GLMC G\,$310.5706\!-\!0.1905$ & A &  13:57:24.3 & -62:04:48 & 0.6 & -131 -- 48 \\
    GLMC G\,$310.6793\!+\!0.4858$ & A &  13:56:51.4 & -61:23:53 & 0.6 & -132 -- 47 \\
    GLMC G\,$311.2947\!-\!0.3779$ & B &  14:03:47.6 & -62:04:09 & 0.6 & -132 -- 32 \\
    GLMC G\,$311.4166\!+\!0.9159$ & A &  14:01:49.3 & -60:47:28 & 0.6 & -132 -- 32 \\
    GLMC G\,$312.3296\!-\!0.0874$ & A &  14:11:27.4 & -61:29:26 & 0.6 & -132 -- 47 \\
    GLMC G\,$312.4354\!-\!0.7425$ & B &  14:14:01.3 & -62:04:51 & 0.6 & -131 -- 47 \\
    GLMC G\,$312.7411\!+\!0.0787$ & A &  14:14:17.5 & -61:12:18 & 0.5 & -132 -- 32 \\
    GLMC G\,$314.4714\!-\!1.0256$ & B &  14:31:03.7 & -61:38:42 & 0.5 & -131 -- 32 \\
    GLMC G\,$315.0424\!+\!0.1836$ & A &  14:31:40.6 & -60:18:35 & 0.5 & -131 -- 32 \\
    GLMC G\,$315.3154\!+\!0.9171$ & B &  14:31:29.4 & -59:31:39 & 0.5 & -131 -- 32 \\
    GLMC G\,$317.4663\!-\!0.0668$ & A &  14:50:09.2 & -59:32:46 & 0.5 & -130 -- 35 \\
    GLMC G\,$317.8897\!-\!0.2531$ & A &  14:53:47.9 & -59:31:25 & 0.6 & -128 -- 51 \\
    GLMC G\,$318.7221\!-\!0.3129$ & A &  14:59:48.4 & -59:11:28 & 0.6 & -127 -- 52 \\
    GLMC G\,$319.3711\!-\!0.7818$ & B &  15:06:01.4 & -59:17:18 & 0.5 & -127 -- 52 \\
    GLMC G\,$321.9359\!-\!0.0061$ & A &  15:19:43.0 & -57:18:07 & 0.5 & -128 -- 51 \\
    GLMC G\,$321.9367\!-\!0.0063$ & A &  15:19:43.3 & -57:18:06 & 0.5 & -128 -- 51 \\
    GLMC G\,$324.8889\!-\!0.4130$ & B &  15:39:06.1 & -55:57:55 & 0.6 & -125 -- 54 \\
    GLMC G\,$326.7527\!+\!0.7101$ & A &  15:44:44.8 & -53:56:44 & 0.6 & -124 -- 55 \\
    GLMC G\,$326.9249\!-\!0.5142$ & A &  15:50:52.9 & -54:48:01 & 0.6 & -124 -- 55 \\
    GLMC G\,$326.9646\!-\!0.0130$ & A &  15:48:56.0 & -54:23:04 & 0.5 & -107 -- 72 \\
    GLMC G\,$326.9872\!-\!0.0306$ & A &  15:49:07.8 & -54:23:02 & 0.7 & -120 -- 50 \\
    GLMC G\,$327.4618\!-\!0.1516$ & B &  15:52:11.0 & -54:10:50 & 0.7 & -120 -- 55 \\
    GLMC G\,$328.0585\!+\!0.3694$ & A &  15:53:04.8 & -53:23:59 & 0.6 & -125 -- 54 \\
    GLMC G\,$328.2162\!-\!0.0216$ & A &  15:55:34.4 & -53:36:02 & 0.6 & -126 -- 40 \\
    GLMC G\,$328.9849\!-\!0.4363$ & B &  16:01:19.1 & -53:25:05 & 0.6 & -125 -- 54 \\
    GLMC G\,$329.1065\!-\!0.0699$ & B &  16:00:19.8 & -53:03:41 & 0.6 & -125 -- 54 \\
    GLMC G\,$329.4935\!+\!0.1053$ & A &  16:01:30.5 & -52:40:32 & 0.5 & -125 -- 54 \\
    GLMC G\,$329.6382\!-\!1.0601$ & B &  16:07:22.8 & -53:27:03 & 0.5 & -125 -- 54 \\
    GLMC G\,$329.7182\!+\!0.8046$ & A &  15:59:37.6 & -51:59:58 & 0.5 & -125 -- 40 \\
    GLMC G\,$331.3051\!-\!0.4970$ & B &  16:12:56.1 & -51:54:22 & 0.7 &  -80 -- 65 \\
    GLMC G\,$331.6119\!-\!0.2711$ & A &  16:13:22.5 & -51:31:52 & 0.8 & -107 -- 72 \\
    GLMC G\,$331.7238\!-\!0.2039$ & A &  16:13:36.1 & -51:24:20 & 0.8 & -107 -- 72 \\
    GLMC G\,$332.1934\!+\!0.1313$ & A &  16:14:18.2 & -50:50:21 & 0.3 & -149 -- 30 \\
    GLMC G\,$334.4183\!+\!0.0581$ & A &  16:24:30.8 & -49:19:52 & 0.5 & -123 -- 42 \\
    GLMC G\,$334.5468\!+\!0.4889$ & A &  16:23:11.4 & -48:56:13 & 0.5 & -123 -- 42 \\ 
    GLMC G\,$336.0800\!+\!0.4807$ & B &  16:29:39.4 & -47:50:32 & 0.2 & -148 -- 31 \\ 
    GLMC G\,$336.2340\!+\!0.5070$ & B & 16:30:10.4 & -47:42:45 & 0.3 & -148 -- 31 \\
    GLMC G\,$338.3874\!+\!0.7252$ & B & 16:37:46.8 & -45:58:56 & 0.7 & -119 -- 60 \\
    GLMC G\,$339.5350\!-\!0.0524$ & B & 16:45:28.7 & -45:38:02 & 0.6 & -121 -- 48 \\
    GLMC G\,$340.0314\!+\!0.6002$ & B & 16:49:25.3 & -45:46:35 & 0.7 & -120 -- 59 \\
    GLMC G\,$340.4215\!-\!0.0222$ & B & 16:44:30.5 & -44:49:55 & 0.3 & -148 -- 31 \\
    GLMC G\,$340.8223\!-\!1.0286$ & B & 16:48:36.7 & -44:56:19 & 0.6 & -119 -- 60 \\
    GLMC G\,$340.9942\!-\!0.3102$ & A & 16:54:27.3 & -45:16:14 & 0.6 & -119 -- 60 \\
    GLMC G\,$341.0242\!+\!0.1975$ & B & 16:51:55.5 & -44:40:59 & 0.5 & -119 -- 47 \\ \hline
  \end{tabular}
  \label{tab:nondet}
\end{table*}    

\begin{table*}
  \contcaption{}
  \begin{tabular}{lcllcc} \hline
    {\bf \glim}    & {\bf Category} & {\bf Right Ascension} & {\bf Declination}
       & {\bf RMS}  & {\bf Velocity Range} \\ 
    {\bf PSC name} &                & {\bf (J2000)}         & {\bf (J2000)}    
       & {\bf (Jy)} & {\bf (\kms)}         \\ [2mm] \hline
    GLMC G\,$342.3190\!+\!0.5876$ & A & 16:49:50.5 & -44:20:08 & 0.7 & -117 -- 62 \\
    GLMC G\,$342.5437\!-\!0.7020$ & B & 16:52:44.7 & -43:05:22 & 0.6 & -117 -- 62 \\
    GLMC G\,$342.7816\!+\!0.1588$ & A & 16:59:03.2 & -43:43:21 & 0.6 & -116 -- 63 \\
    GLMC G\,$342.8209\!+\!0.3827$ & A & 16:56:09.8 & -43:00:01 & 0.6 & -119 -- 60 \\
    GLMC G\,$343.7378\!-\!0.1116$ & A & 16:55:20.7 & -42:49:46 & 0.5 & -118 -- 61 \\
    GLMC G\,$344.1581\!-\!0.0748$ & A & 17:00:33.0 & -42:25:07 & 0.6 & -118 -- 47 \\
    GLMC G\,$344.2931\!-\!0.7158$ & A & 17:01:47.3 & -42:03:52 & 0.6 & -118 -- 47 \\
    GLMC G\,$345.2574\!-\!0.0351$ & A & 17:04:58.6 & -42:20:53 & 0.3 & -149 -- 30 \\
    GLMC G\,$345.6528\!-\!1.0043$ & B & 17:10:35.7 & -41:25:57 & 0.7 & -115 -- 64 \\
    GLMC G\,$345.7172\!+\!0.8166$ & B & 17:03:06.4 & -40:17:09 & 0.4 & -145 -- 34 \\
    GLMC G\,$345.8590\!-\!0.6792$ & B & 17:09:51.3 & -41:04:25 & 0.3 & -149 -- 30 \\
    GLMC G\,$346.1045\!-\!0.4612$ & B & 17:09:42.0 & -40:44:47 & 0.3 & -148 -- 30 \\
    GLMC G\,$347.2918\!+\!0.1330$ & A & 17:10:53.1 & -39:26:18 & 0.4 & -145 -- 34 \\
    GLMC G\,$347.4007\!+\!0.1229$ & A & 17:11:15.7 & -39:21:24 & 0.4 & -145 -- 34 \\
    GLMC G\,$347.7014\!+\!0.9183$ & B & 17:08:53.2 & -38:38:32 & 0.4 & -144 -- 34 \\
    GLMC G\,$348.5661\!+\!0.1626$ & A & 17:14:36.9 & -38:23:25 & 0.5 & -113 -- 66 \\
    GLMC G\,$348.9414\!-\!0.1082$ & B & 17:16:51.1 & -38:14:33 & 0.6 & -112 -- 67 \\
    GLMC G\,$349.1465\!-\!0.9766$ & B & 17:21:04.9 & -38:34:25 & 0.5 & -112 -- 67 \\
    GLMC G\,$349.2867\!+\!0.4602$ & A & 17:15:31.2 & -37:37:53 & 0.5 & -112 -- 67 \\
    GLMC G\,$349.3072\!+\!0.3703$ & B & 17:15:57.0 & -37:40:01 & 0.5 & -112 -- 67 \\
    GLMC G\,$349.5204\!-\!0.2641$ & B & 17:19:11.6 & -37:51:34 & 0.5 & -112 -- 67 \\
    GLMC G\,$349.8799\!+\!0.5500$ & A & 17:16:53.1 & -37:05:47 & 0.6 & -112 -- 67 \\
    GLMC G\,$011.1080\!+\!0.2045$ & B & 18:09:17.1 & -19:13:19 & 0.6 & -56  -- 123 \\
    GLMC G\,$011.1158\!+\!0.0513$ & B & 18:09:52.2 & -19:17:21 & 0.6 & -56  -- 123 \\
    GLMC G\,$011.2354\!-\!0.9595$ & B & 18:13:52.4 & -19:40:14 & 0.6 & -56  -- 123 \\
    GLMC G\,$011.3446\!+\!0.7971$ & A & 18:07:34.8 & -18:43:39 & 0.6 & -56  -- 123 \\
    GLMC G\,$011.3802\!-\!0.0068$ & B & 18:10:37.5 & -19:05:09 & 0.5 & -56  -- 123 \\
    GLMC G\,$012.2200\!-\!0.3339$ & A & 18:13:32.6 & -18:30:24 & 0.6 & -56  -- 123 \\
    GLMC G\,$012.5764\!+\!0.3089$ & B & 18:11:53.3 & -17:53:08 & 0.2 & -58  -- 121 \\
    GLMC G\,$012.6896\!+\!0.2881$ & B & 18:12:11.7 & -17:47:46 & 0.5 & -56  -- 123 \\
    GLMC G\,$013.2348\!-\!0.0723$ & A & 18:14:37.1 & -17:29:25 & 0.2 & -58  -- 120 \\
    GLMC G\,$013.2690\!-\!0.3487$ & A & 18:15:42.4 & -17:35:32 & 0.7 & -59  -- 119 \\
    GLMC G\,$013.2853\!+\!0.5723$ & B & 18:12:21.0 & -17:08:14 & 0.5 & -56  -- 123 \\
    GLMC G\,$013.2894\!+\!0.2129$ & B & 18:13:40.7 & -17:18:21 & 0.5 & -56  -- 123 \\
    GLMC G\,$014.1087\!-\!0.5688$ & A & 18:18:11.3 & -16:57:28 & 0.5 & -55  -- 124 \\
    GLMC G\,$014.1520\!-\!0.5173$ & B & 18:18:05.1 & -16:53:43 & 0.6 & -55  -- 124 \\
    GLMC G\,$014.4758\!-\!0.1259$ & A & 18:17:17.2 & -16:25:28 & 0.6 & -55  -- 124 \\
    GLMC G\,$014.6786\!+\!0.0394$ & A & 18:17:04.9 & -16:10:03 & 0.2 & -58  -- 120 \\
    GLMC G\,$014.7210\!+\!0.1682$ & B & 18:16:41.6 & -16:04:09 & 0.3 & -58  -- 120 \\
    GLMC G\,$014.8008\!-\!0.1372$ & A & 18:17:58.3 & -16:08:38 & 0.3 & -59  -- 120 \\
    GLMC G\,$015.0296\!+\!0.8534$ & A & 18:14:48.1 & -15:28:17 & 0.3 & -59  -- 120 \\ 
    GLMC G\,$015.0918\!-\!0.2690$ & A & 18:19:01.7 & -15:57:00 & 0.6 & -51 -- 120 \\
    GLMC G\,$015.2571\!-\!0.1561$ & A & 18:18:56.3 & -15:45:03 & 0.6 & -57 -- 121 \\
    GLMC G\,$015.5709\!-\!0.7953$ & B & 18:21:53.8 & -15:46:32 & 0.6 & -57 -- 122 \\ 
    GLMC G\,$016.0634\!-\!0.9387$ & B & 18:23:22.9 & -15:24:29 & 0.5 & -57 -- 122 \\
    GLMC G\,$016.4464\!-\!0.3874$ & A & 18:22:06.6 & -14:48:40 & 0.5 & -57 -- 122 \\
    GLMC G\,$016.6638\!+\!0.6325$ & B & 18:18:49.1 & -14:08:18 & 0.5 & -57 -- 122 \\
    GLMC G\,$016.6786\!+\!0.1714$ & B & 18:20:31.4 & -14:20:35 & 0.7 & -55 -- 124 \\
    GLMC G\,$016.8016\!-\!0.2707$ & A & 18:22:22.4 & -14:26:33 & 0.7 & -55 -- 124 \\
    GLMC G\,$016.8056\!+\!0.8149$ & B & 18:18:25.9 & -13:55:38 & 0.7 & -55 -- 124 \\
    GLMC G\,$016.9750\!-\!0.0291$ & A & 18:21:49.8 & -14:10:34 & 0.7 & -55 -- 124 \\
    GLMC G\,$017.1767\!-\!0.8367$ & B & 18:25:09.9 & -14:22:35 & 0.3 & -59 -- 119 \\
    GLMC G\,$017.5043\!-\!0.6885$ & B & 18:25:15.3 & -14:01:03 & 0.6 & -54 -- 125 \\
    GLMC G\,$018.0544\!+\!0.9845$ & B & 18:20:14.5 & -12:44:47 & 0.6 & -54 -- 125 \\
    GLMC G\,$018.4478\!-\!0.3116$ & B & 18:25:41.8 & -13:00:26 & 0.7 & -54 -- 125 \\
    GLMC G\,$019.0529\!+\!0.6390$ & B & 18:23:24.8 & -12:01:40 & 0.7 & -54 -- 125 \\
    GLMC G\,$019.7592\!+\!0.1095$ & A & 18:26:40.5 & -11:39:02 & 0.6 & -56 -- 123 \\
    GLMC G\,$019.8905\!+\!0.2470$ & B & 18:26:25.7 & -11:28:14 & 0.6 & -56 -- 123 \\
    GLMC G\,$019.9230\!-\!0.2581$ & B & 18:28:18.9 & -11:40:37 & 0.6 & -55 -- 124 \\
    GLMC G\,$020.5655\!+\!0.9478$ & B & 18:25:11.1 & -10:32:48 & 0.6 & -55 -- 123 \\ \hline
    \end{tabular}
\end{table*}

\begin{table*}
  \contcaption{}
  \begin{tabular}{lcllcc} \hline
    {\bf \glim}    & {\bf Category} & {\bf Right Ascension} & {\bf Declination}
       & {\bf RMS}  & {\bf Velocity Range} \\ 
    {\bf PSC name} &                & {\bf (J2000)}         & {\bf (J2000)}    
       & {\bf (Jy)} & {\bf (\kms)}         \\ [2mm] \hline
    GLMC G\,$020.8544\!+\!0.4858$ & B & 18:27:23.8 & -10:30:23 & 0.5 & -55 -- 124 \\
    GLMC G\,$021.5349\!+\!1.0505$ & A & 18:26:39.2 & -09:38:28 & 0.6 & -55 -- 124 \\
    GLMC G\,$022.1572\!-\!0.5487$ & A & 18:33:34.5 & -09:49:51 & 0.8 & -52 -- 127 \\
    GLMC G\,$022.2912\!+\!0.8678$ & B & 18:28:44.0 & -09:03:24 & 0.7 & -53 -- 126 \\
    GLMC G\,$022.4514\!-\!0.4332$ & B & 18:33:42.5 & -09:31:00 & 0.7 & -52 -- 127 \\
    GLMC G\,$022.9242\!+\!0.2235$ & B & 18:32:14.0 & -08:47:39 & 0.7 & -52 -- 127 \\
    GLMC G\,$023.4058\!+\!0.4485$ & A & 18:32:19.5 & -08:15:47 & 0.7 & -52 -- 127 \\
    GLMC G\,$023.4310\!-\!0.5183$ & A & 18:35:50.5 & -08:41:11 & 0.8 & -51 -- 128 \\
    GLMC G\,$023.6155\!-\!0.0071$ & A & 18:34:21.0 & -08:17:14 & 0.8 & -51 -- 128 \\
    GLMC G\,$024.3718\!+\!0.2933$ & A & 18:34:40.8 & -07:28:39 & 0.7 & -52 -- 127 \\
    GLMC G\,$024.6333\!+\!0.1532$ & A & 18:35:40.1 & -07:18:35 & 0.6 & -53 -- 126 \\
    GLMC G\,$025.9114\!-\!1.0679$ & B & 18:42:24.2 & -06:44:02 & 0.6 & -53 -- 126 \\
    GLMC G\,$026.4958\!+\!0.7106$ & B & 18:37:07.3 & -05:23:58 & 0.6 & -53 -- 126 \\
    GLMC G\,$026.5027\!-\!0.5410$ & B & 18:41:36.3 & -05:58:02 & 0.5 & -52 -- 126 \\
    GLMC G\,$027.1474\!-\!0.1431$ & A & 18:41:22.1 & -05:12:43 & 0.6 & -50 -- 129 \\
    GLMC G\,$027.6673\!-\!0.0548$ & B & 18:42:00.5 & -04:42:33 & 0.7 & -50 -- 129 \\
    GLMC G\,$028.0827\!-\!0.7427$ & B & 18:45:13.6 & -04:39:14 & 0.7 & -50 -- 129 \\
    GLMC G\,$028.8128\!+\!0.2641$ & A & 18:42:58.3 & -03:32:41 & 0.6 & -52 -- 127 \\
    GLMC G\,$029.3077\!+\!0.3218$ & A & 18:43:40.4 & -03:04:41 & 0.6 & -52 -- 127 \\
    GLMC G\,$030.1029\!-\!0.0789$ & A & 18:46:33.3 & -02:33:13 & 0.7 & -49 -- 130 \\
    GLMC G\,$030.3943\!-\!0.7064$ & B & 18:49:19.4 & -02:34:50 & 0.7 & -48 -- 130 \\
    GLMC G\,$030.4003\!-\!1.0657$ & B & 18:50:36.9 & -02:44:20 & 0.7 & -48 -- 131 \\
    GLMC G\,$030.6039\!+\!0.1760$ & A & 18:46:33.7 & -01:59:29 & 0.7 & -49 -- 130 \\
    GLMC G\,$030.6670\!-\!0.3319$ & B & 18:48:29.2 & -02:10:01 & 0.7 & -49 -- 130 \\
    GLMC G\,$030.6874\!+\!0.3592$ & B & 18:46:03.7 & -01:50:01 & 0.6 & -51 -- 128 \\
    GLMC G\,$031.6615\!+\!0.3668$ & B & 18:47:48.8 &  00:57:48 & 0.6 & -48 -- 131 \\
    GLMC G\,$033.5681\!-\!0.3841$ & B & 18:53:57.9 &  00:23:28 & 0.6 & -48 -- 131 \\
    GLMC G\,$034.4049\!+\!0.0316$ & B & 18:54:00.7 &  01:19:31 & 0.7 & -48 -- 131 \\
    GLMC G\,$034.6817\!+\!0.8551$ & A & 18:51:35.0 &  01:56:50 & 0.6 & -50 -- 129 \\ 
    GLMC G\,$036.1695\!+\!0.2392$ & B & 18:56:29.7 &  02:59:25 & 0.6 & -49 -- 130 \\
    GLMC G\,$037.2004\!-\!0.0019$ & B & 18:59:14.4 &  03:47:50 & 0.6 & -48 -- 131 \\
    GLMC G\,$038.9985\!-\!0.5075$ & A & 19:04:20.8 &  05:09:52 & 0.7 & -45 -- 134 \\
    GLMC G\,$040.1332\!+\!0.9389$ & B & 19:01:16.0 &  06:50:08 & 0.6 & -45 -- 134 \\
    GLMC G\,$040.1647\!+\!0.2958$ & B & 19:03:37.6 &  06:34:09 & 0.6 & -45 -- 134 \\
    GLMC G\,$041.2075\!-\!0.0235$ & B & 19:06:41.9 &  07:20:57 & 0.7 & -45 -- 134 \\
    GLMC G\,$041.9910\!+\!0.0795$ & B & 19:07:47.1 &  08:05:31 & 0.7 & -45 -- 134 \\
    GLMC G\,$042.0982\!+\!0.3515$ & B & 19:07:00.5 &  08:18:45 & 0.5 & -47 -- 132 \\
    GLMC G\,$042.1608\!+\!0.6079$ & A & 19:06:12.2 &  08:29:09 & 0.4 & -57 -- 123 \\
    GLMC G\,$042.8347\!-\!0.2667$ & B & 19:10:36.0 &  08:40:51 & 0.6 & -44 -- 134 \\
    GLMC G\,$043.1195\!+\!0.3682$ & B & 19:08:51.1 &  09:13:36 & 0.6 & -45 -- 134 \\
    GLMC G\,$044.2406\!+\!0.3086$ & A & 19:11:10.0 &  10:11:37 & 0.7 & -44 -- 135 \\
    GLMC G\,$045.6999\!-\!0.2536$ & B & 19:15:57.0 &  11:13:32 & 0.4 & -58 -- 122 \\
    GLMC G\,$046.4995\!+\!0.8093$ & A & 19:13:37.4 &  12:25:39 & 0.6 & -46 -- 133 \\
    GLMC G\,$047.4102\!+\!0.1366$ & B & 19:17:48.1 &  12:55:13 & 0.6 & -45 -- 134 \\
    GLMC G\,$048.6717\!-\!0.3044$ & A & 19:21:50.2 &  13:49:39 & 0.6 & -44 -- 135 \\
    GLMC G\,$050.0366\!+\!0.0714$ & A & 19:23:07.6 &  15:12:30 & 0.6 & -44 -- 135 \\
    GLMC G\,$050.8533\!+\!0.3305$ & A & 19:23:46.8 &  16:03:02 & 0.6 & -44 -- 135 \\
    GLMC G\,$052.0790\!+\!0.2581$ & A & 19:26:28.4 &  17:05:43 & 0.6 & -43 -- 135 \\
    GLMC G\,$052.5815\!+\!0.2014$ & A & 19:27:41.1 &  17:30:36 & 0.7 & -41 -- 138 \\
    GLMC G\,$059.1885\!+\!0.1053$ & B & 19:41:44.7 &  23:14:19 & 0.7 & -40 -- 139 \\
    GLMC G\,$059.4914\!-\!0.0831$ & B & 19:43:06.8 &  23:24:29 & 0.7 & -40 -- 139 \\
    GLMC G\,$062.8607\!+\!0.8219$ & B & 19:47:06.1 &  26:46:40 & 0.8 & -39 -- 140 \\
    GLMC G\,$062.9230\!-\!0.5845$ & B & 19:52:39.3 &  26:07:00 & 0.7 & -39 -- 140 \\ \hline
    \end{tabular}
\end{table*}    

Figure~\ref{fig:detections} shows Hanning smoothed spectra (velocity
resolution 0.088~km/s) of each of the detected 6.7-GHz methanol masers
made with the Hobart radio telescope.  In many cases the only spectra
available in the literature for the previously detected sources were
observed more than a decade ago \citep{CVEWN95,WHRB97} and so the
spectra of all detected sources are shown here to enable variability
comparisons to be made.  The sources which show significant
variability are noted in section~\ref{sec:indiv}.

The last column in Table~\ref{tab:det} shows the separation between
the methanol maser and the {\em GLIMPSE} source.  As expected from the
process used to select the target {\em GLIMPSE} sources, all the
previously known methanol masers are separated from them by more than
3.5~arcmin.  In general the newly discovered sources are within a few
arcminutes of the {\em GLIMPSE} source, although in most cases the
separation is significantly greater than the pointing accuracy
(0.5~arcmin), so it is unlikely that the masers and infrared sources
are directly associated.

\begin{figure*}
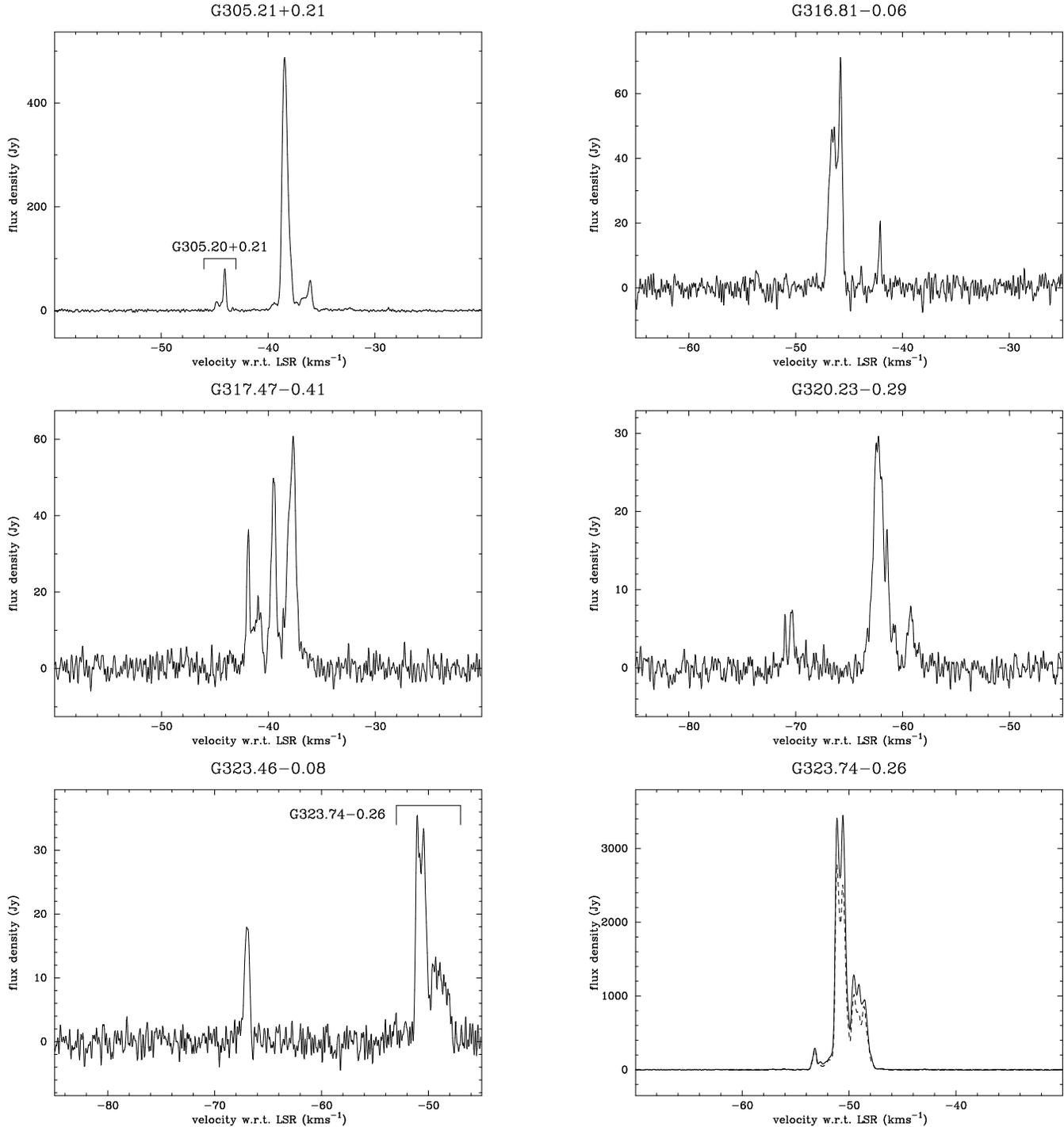

  \specdfig{figure2a}{figure2b}
  \specdfig{figure2c}{figure2d}
  \specdfig{figure2e}{figure2f}
  \caption{Spectra of the 6.7-GHz methanol maser sources detected in
    the \glim-based search.  The dashed spectra shown for
    G323.740-0.262 is taken from \citet{ECL+04}}
  \label{fig:detections}
\end{figure*}

\begin{figure*}
  \specdfig{figure2g}{figure2h}
  \specdfig{figure2i}{figure2j}
  \specdfig{figure2k}{figure2l}
  \contcaption{The dashed spectra shown for G326.66+0.52, G328.24-0.55
    and G331.56-0.12 are taken from \citet{EVM+96} and those for
    G337.15-0.40 and G337.39-0.20 are taken from \citet{E96}.}
\end{figure*}

\begin{figure*}
  \specdfig{figure2m}{figure2n}
  \specdfig{figure2o}{figure2p}
  \specdfig{figure2q}{figure2r}
  \contcaption{NOTE: The spectrum of G345.20-0.04 was observed at
    Ceduna and has a velocity resolution of 0.11~\kms, compared to
    0.09~\kms\/ for the other spectra.}
\end{figure*}

\begin{figure*}
  \specdfig{figure2s}{figure2t}
  \specdfig{figure2u}{figure2v}
  \specdfig{figure2w}{figure2x}
  \contcaption{The dashed spectra shown for G25.82-0.17 and G29.86-0.04 are
    taken from \citet{E96}.}
\end{figure*}

\begin{figure*}
  \specdfig{figure2y}{figure2z}
  \specdfig{figure2aa}{figure2ab}
  \specdfig{figure2ac}{figure2ad}
  \contcaption{}
\end{figure*}

\begin{figure*}
  \specsfig{figure2ae}
  \contcaption{}
\end{figure*}

\subsection{Comments on individual sources} \label{sec:indiv}

{\em G305.20+0.21/G305.21+0.21/G305.25+0.25:} G305.20+0.21 and
G305.21+0.21 are separated by 22~arcsec and hence cannot be readily
distinguished by these observations.  \citet{WB06} suggest that this
pair of sources consists of two star formation regions in different
evolutionary stages, G305.21+0.21 being younger and G305.20+0.21 more
evolved.  G305.21+0.21 has a very similar peak intensity and spectral
morphology to that seen more than a decade ago by \citet{CVEWN95} and
\citet*{GGV04} observed this source to show relatively little
variability.  G305.20+0.21 shows a modest increase in intensity, also
consistent with the observations of \citet{GGV04}

Emission from G305.25+0.25 was also detected in this observation
covering a velocity range from -33 -- -28~\kms, but is too weak to be
easily seen on the scale shown in Fig.~\ref{fig:detections}.  These
observations show the source to have peaks near -29 and -32~\kms, each
of about 6~Jy, significantly stronger than the only previous
observation of G305.25+0.25 by \citet{CVEWN95} in September 1992, who
observed it to have a peak flux density of about 2 Jy.

{\em G316.81-0.06:} This source has been observed on numerous previous
occasions; the observations of \citet{MG92} measured the flux density of
the emission at -46~\kms\/ to be about 10~Jy in late 1991.  The
observations of \citet*{CVE95} in 1992/3 exhibit significant changes in
the relative flux density of the different spectral features, and they
list it as showing significant variability.
Figure~\ref{fig:detections} shows that two strongest spectral
components in this source have increased in flux density by more than a
factor of four since 1992, while some of the weaker features have
dropped below the detection limit of the current observations.

{\em G317.47-0.41:} This is the strongest of the newly detected 6.7-GHz
methanol masers with a peak flux density of approximately 60~Jy and
three strong peaks over a velocity range of 6~\kms.

{\em G320.23-0.29:} Although this source is listed as being variable
by \citet{CVE95} the velocity range, peak flux density and general
appearance are very similar to those observed more than a decade ago
\citep{MG92,CVEWN95,WHRB97}.  The only exception is that a spectral
feature present near -69.5~\kms\/ with a peak flux density of about
4 Jy is not present in the current observations.


{\em G323.74-0.26:} This is one of the strongest 6.7-GHz methanol
masers and has been the subject of detailed study by \citet{WLB02}.
Its intensity is also the reason that it was found by this survey, as
it was detected in observations of a {\em GLIMPSE} source more than 20
arcmin from the maser site.  It is listed as not variable by
\citet{CVEWN95}, but \citet{GGV04} observed the source to exhibit
moderate variations, some highly correlated between spectral features.
A comparison of the spectrum in Fig.~\ref{fig:detections} with that of
\citet{CVEWN95} shows an increase of the order of 10 per cent in the
peak flux density and changes in the relative intensity of most
features.  The 6.7-GHz methanol maser emission in this source was
observed in late 2003 (two years prior to the current observations) by
\citet{ECL+04} and that spectrum is shown using a dashed line in
Fig.~\ref{fig:detections}.  The velocity range reported in
Table~\ref{tab:maserprop} is slightly larger than reported in previous
observations \citep{CVEWN95,PNE+98}, extending to a maximum velocity
of -42.8~\kms\/ (compared to -46~\kms).

{\em G326.66+0.52:} The only previous observations of this source were
made by \citet{EVM+96} and Fig.~\ref{fig:detections} shows that it has
changed dramatically in the decade since then, with a number of new
spectral features present and an increase in peak flux density of
approximately 100 per cent.  The nearby source G326.64+0.61, which
partially overlaps the velocity range of G326.66+0.52 has also
increased significantly in peak flux density since the 1994
observations of \citet{EVM+96}.

{\em G328.24-0.55/G328.25-0.53:} This pair of sources is separated by
83~arcsec and was detected in the two largest surveys of southern
6.7-GHz methanol masers \citep{CVEWN95,WHRB97}.  It has also been
studied at high resolution by \citet{PNE+98} and \citet{DOE04} who
found both sites of maser emission to be offset from nearby radio
continuum emission.  Somewhat unusually for a moderately strong
methanol maser, this source has demonstrated significant variablity in
the strongest spectral features.  \citet{GGV04} monitored this source
for 4 years and their observations show that the flux density of the
emission at -44.7~\kms\/ monotonically increased from around 550 Jy in
1999 to 800~Jy in 2003.  Figure~\ref{fig:detections} compares two
observations of this source separated by a decade (the spectrum from
\cite{EVM+96} uses a dashed line), and shows that the increase in the
strongest emission feature has continued and in late 2005 it had a
peak flux density of nearly 1000 Jy.

{\em G331.56-0.12:} This source is listed as not variable by
\citet{CVEWN95}, but comparison of the observations in this paper with
those of \citet{EVM+96} shows significant variabilty is present on
longer timescales.

{\em G337.15-0.40:} This source was first detected by \citet{E96}, but
has not been reported in any refereed publication and so is listed as
a new detection here.  It's in the Galactic longitude range covered by
\citet{C96}, but too far from the Galactic plane to be detected in
that search.  The overall spectral shape has changed relatively little
over the decade between the two observations shown in
Fig.~\ref{fig:detections}.

{\em G337.39-0.20:} Similar to G337.15-0.40 this source was first
detected by \citet{E96}, but has not been reported in any refereed
publication.  The earlier observations show that the velocity range is
larger than is apparent from the present epoch and that the peak flux
density has decreased by nearly 50 per cent in the decade between 1995
and 2005.

{\em G339.88-1.26:} This is one of the strongest 6.7-GHz methanol
masers known, and it was detected in a distant sidelobe of an
observation of a {\em GLIMPSE} source more than half a degree away.
This source has been the focus of a number of detailed studies as the
masers have a linear spatial distribution along the same position
angle as extended mid-infrared emission and are projected against a
weak ultra-compact HII region
\citep*{ENM96,NWC+93,DWP+02}. \citet{GGV04} monitored this source for
4 years and found it to show relatively little variation in most
spectral features.  Figure~\ref{fig:detections} shows its peak flux
density has decreased slightly in comparison to that observed by
\citet{CVEWN95} and \citet{WHRB97} and several of the weaker features
near 30~\kms\/ observed by \citet{CVEWN95} are no longer detectable.

{\em G340.97-1.03:} This new detection shows emission over a velocity
range of more than 10~\kms\/, separated into three distinct groups.
Most of the spectral features have a flux density of a few Jy, with
the strongest at about 10~Jy.  The measured position of the maser is
offset by 3.5~arcmin from the {\em GLIMPSE} source and so the two are
very unlikely to be associated.

{\em G343.52-0.50:} This is one of the weakest 6.7-GHz methanol masers
detected in this survey and although the signal to noise ratio of the
spectrum shown in Fig.~\ref{fig:detections} is low it was detected on
a total of three separate occasions.  The measured position of the maser
is 1.7~arcmin from the {\em GLIMPSE} source and so it is possible, but
not likely that the two are associated.

{\em G345.20-0.04:} This is another newly discovered source and is the
only one for which the final observation was made at Ceduna rather
than Hobart.  As a consequence the velocity resolution of the spectrum
is 0.105~\kms, compared to 0.088~\kms\/ for the other sources.  The
measured position of the maser is offset by more than 3.5~arcmin from
the {\em GLIMPSE} source and so the two are very unlikely to be
associated.

{\em G11.15-0.14:} This moderately strong newly discovered maser has
emission covering nearly 10~\kms\/.  As for many of the other new
discoveries, the maser emission is significantly offset (6.2~arcmin in
this case) from the {\em GLIMPSE} source towards which the search was
made.

{\em G18.99-0.04:} This newly discovered 6.7-GHz methanol maser is
offset by 1.5 arcmin from the {\em GLIMPSE} source toward which the
search was made, so it is possible, but not likely that the two are
associated.

{\em G23.01-0.41:} The first observation of this strong 6.7-GHz
methanol maser was in the 6.7-GHz methanol maser discovery paper of
\citet{M91}.  \citet{CVEWN95} lists this source as being slightly
variable, however, comparison of the spectrum in
Fig.~\ref{fig:detections} with those taken more than a decade earlier
shows the peak flux density and overall shape of the spectrum have
changed very little, although a weak spectral feature near 70~\kms\/
is no longer present.

{\em G24.14+0.00:} This source was discovered by \citet{SK00} in 1999,
and they found it to have a peak flux density of 40~Jy, while the
observations of \citet{SKHKP02} made approximately a year later list a
lower value of 26~Jy.  Figure~\ref{fig:detections} shows that 5 years
later its peak flux density lies between the two previously reported
values and that it continues to show emission over a narrow velocity
range.

{\em G25.82-0.17:} This source was first discovered by \citet{SVGM93}
and has subsequently been detected in a number of {\em IRAS}-based
\citep{WHRB97,SK00} and untargeted surveys \citep{E96,SKHKP02}.  The
peak flux density in the spectrum of \citet{SVGM93} is similar to that
seen in Fig.~\ref{fig:detections} (although it doesn't appear to agree
with the value listed in Table~2 of \citeauthor{SVGM93}), while
\citet{SKHKP02} report 66~Jy.  The observations of \citet{SK00} show a
peak flux density of approximately 20~Jy, but it appears that this is
largely due to an offset between the maser location and the position
of the {\em IRAS} source targeted.  Figure~\ref{fig:detections} shows
the spectrum observed by \citet{E96} as a dashed line, which along
with the other observations suggests that the source shows variations
of tens of per cent on timescales of years.

{\em G27.21+0.26:} This source was discovered by \citet{SKHKP02} and
has roughly doubled in peak flux density in the 5 years since their
observations.

{\em G27.28+0.15:} This source was discovered by \citet{SVK+99} who
quote its peak flux density as 41~Jy in 1995, although their spectrum
shows approximately 28~Jy.  It has also been observed by \citet{SK00}
and \citet{SKHKP02} who measured its peak flux density to be 21 and
23~Jy respectively.  Figure~\ref{fig:detections} shows the current peak
flux density to have increased to 36~Jy.

{\em G28.02-0.44:} This source was discovered by \citet{SKHKP02} who
observed emission of comparable strength over the same, large velocity
range as seen in Fig.~\ref{fig:detections}.

{\em G29.86-0.04:} This source has been previously observed by a
number of other authors and shows modest variability.  Its peak flux
density was 67~Jy in 1992 \citep{CVEWN95}, 57~Jy in 1995 \citep{E96},
52~Jy in 2000 \citep{SKHKP02}, but is currently in excess of 80 Jy.
Figure~\ref{fig:detections} shows the spectrum observed by \citet{E96}
as a dashed line and significant variations in the relative intensity
are clearly apparent.

{\em G30.20-0.17:} This source was detected at the edge of the
observed velocity range.  Comparison of the spectrum in
Fig.~\ref{fig:detections} with that from \citet{CVEWN95} shows
emission previously extended to lower velocities and this may still be
present, but beneath the detection threshold of the current
observations.  The peak and integrated flux densities are comparable
to those reported by \citet{SKHKP02}.

{\em G30.70-0.07:} This source was first observed by \citet{M91}, who
reported a peak flux density of 125~Jy in 1990.  \citet{CVEWN95}
measured a lower peak flux density of 88~Jy two years later and list it
as being a slightly variable source.  The current observations show a
greater peak flux density than any previously reported (154~Jy),
although the strength of some of the secondary features has not changed
as much.

{\em G30.79+0.20:} This source was discovered by \citet{CVEWN95} who
found it to exhibit emission in two groups over a velocity range of
75--89~\kms, similar to that observed by \cite{SKHKP02}.
Figure~\ref{fig:detections} shows emission over this range, but also a
number of weak features near 100~\kms.  Given that emission in this
range was not observed on previous occasions, it is probably the
nearby source G30.89+0.17 detected in the telescope sidelobes.  The
peak flux density of the emission near 76~\kms\/ has more than tripled
in the period between 1992 and 2005.

{\em G30.82-0.05:} This source shows emission over an usually large
velocity range (nearly 20~\kms).  In previous observations the peak
flux density has been observed near 101~\kms, however, this spectral
feature has gradually declined in strength over the last 15 years.
The strongest spectral feature in the current observations is at a
velocity of 91.7~\kms.

{\em G31.04+0.36:} This source was discovered by \citet{SK00} who
observed it to have a slightly greater flux density, but the spectrum
is overall very similar to that shown in Fig.~\ref{fig:detections}.

{\em G35.18-0.74:} This source was discovered when making a 5-point
grid observation of G35.20-0.74 which showed the highest velocity
emission is offset by 1.3 arcmin to the southwest.  Comparison of the
spectra of G35.18-0.74 and G35.20-0.74 in Fig.~\ref{fig:detections}
shows that their velocity ranges partially overlap, although
G35.18-0.74 has a number of spectral features which are not blended,
with the strongest emission at velocities greater than 34~\kms.
Emission from this source is not evident in the G35.20-0.74 spectrum
of \citet{CVEWN95}, but can be seen in the {\em IRAS}18556+0136
spectrum of \citet{SK00}, although it was not recognised as a separate
source.

{\em G35.20-0.74/G35.18-0.74:} \citet{CVEWN95} did not detect any
variability in G35.20-0.74 in their 1992/3 observations, however it
has increased in peak flux density by approximately 20 per cent over
the last 13 years.  Emission from G35.20-0.74 extends to about
35~\kms, though this feature overlaps in velocity with the strongest
emission in G35.18-0.74.

\section{Discussion} \label{sec:discussion}

The primary purpose of the observations presented in this paper was to
determine whether, of the {\em GLIMPSE} sources identified by E06 as
being likely to be young high-mass star formation regions with an
associated methanol maser, those with the most extreme 8.0-\micron\/
intensity, or reddest [3.6]-[4.5] colours have a greater detection
rate than the average.  E06 predicted that around 11 per cent of {\em
  GLIMPSE} sources meeting the criteria of [8.0] $<$ 10 magnitude and
[3.6]-[4.5] $>$ 1.3 should have an associated methanol maser within
3.5~arcmin.  A search towards 200 {\em GLIMPSE} sources meeting these
criteria should therefore yield around 22 methanol maser detections
within 3.5~arcmin of the infrared source.  A total of thirty-eight
6.7-GHz methanol masers were detected in observations towards 27 of
the {\em GLIMPSE} sources, however, only nine of these are new
discoveries.  For the other nineteen {\em GLIMPSE} sources where a
previously detected 6.7-GHz methanol maser source was observed, in all
cases the separation is more than 3.5~arcmin.  Of the nine new
discoveries, six are within 3.5~arcmin of the {\em GLIMPSE} source,
but only one (G317.47-0.41) is at a separation of less than 1~arcmin.
This means that the number of newly detected masers associated with
{\em GLIMPSE} sources is less than 30 per cent of the number predicted
(22).  These results show that the hypothesis that the most extreme
{\em GLIMPSE} sources meeting the E06 criteria are more likely to have
an associated 6.7-GHz methanol maser is incorrect and the reasons why
need to be investigated.

The methanol maser selection criteria of E06 only set a lower bound
for the [3.6]-[4.5] colour and an upper bound for the 8.0-\micron\/
intensity.  It may be the case that the {\em GLIMPSE} point sources
associated with 6.7-GHz methanol masers are confined to a narrower
range of 8.0 micron intensities and/or [3.6]-[4.5] colours within this
box, and lower and upper bounds respectively should be sought.  The
E06 criteria were determined by examining the {\em GLIMPSE} point
sources associated with a sample of 189 6.7-GHz methanol masers drawn
from \citet{C96}, \citet{WBHR98} and \citet{E05}, whose positions are
all known to sub-arcsecond accuracy.  Of these 189 6.7-GHz methanol
masers 82 have a {\em GLIMPSE} point source within 2 arcseonds.  Sixty
three of the {\em GLIMPSE} point sources with an associated methanol
maser have an 8.0-\micron\/ intensity measurement and for seven (11
per cent) it is less than 4.58 magnitudes.  Fifty three of the {\em
  GLIMPSE} point sources have both 3.6- and 4.5-\micron\/
measurements, and of these the [3.6]-[4.5] colour is greater than 2.87
magnitudes for eight (15 per cent).  So only a relatively small
percentage of currently known 6.7-GHz methanol maser sources match the
more extreme criteria used here to select target {\em GLIMPSE} point
sources.  However, the more important question is whether this is
because only a small number of sources with these criteria are
associated with methanol masers, or if it is because there are
relatively few sources that meet these criteria.  The 2005 April 15
release of the {\em GLIMPSE} point source catalogue contains 30.2
million sources and of these there are 138 which meet the criteria
8.0-\micron\/ intensity less than 10.0 magnitudes and [3.6]-[4.5]
colour $>$ 2.87 magnitudes (category A) and 122 which meet the
criteria 8.0-\micron\/ intensity less than 4.58 magnitudes and
[3.6]-[4.5] colour $>$ 1.3 magnitudes (category B).  A total of 5675
{\em GLIMPSE} point sources meet the E06 criteria, of which
approximately two per cent fall within the category A and category B
regions.  So approximately a quarter of the 6.7-GHz methanol masers
for which a {\em GLIMPSE} association can reliably determined are
associated with sources in either category A or B, which represent
less than 5 per cent of the sources meeting the E06 criteria.  So the
{\em a priori} evidence was that the most extreme sources meeting the
E06 criteria do have a greater than average chance of having an
associated 6.7-GHz methanol maser.

If 6.7-GHz methanol masers associated with {\em GLIMPSE} point sources
in category A and/or B typically have a high peak flux density then
they would be more likely to have been detected in previous searches.
Of the eight known 6.7-GHz methanol masers associated with category A
sources, six have a peak flux density greater than 30~Jy.  In contrast
less than 30 per cent of the 519 sources in the \citeauthor{PMB05}
catalogue of 6.7-GHz methanol masers have a peak flux density greater
than 30~Jy.  For the seven 6.7-GHz methanol masers associated with a
category B source, three have a peak flux density greater than 30~Jy.
Given the small number of masers associated with {\em GLIMPSE} point
sources in either category A or B, it isn't possible to draw firm
conclusions, but it appears likely that those associated with category
A sources often have a higher peak flux density than the majority of
6.7-GHz methanol masers.  This does not necessarily imply that these
masers are intrinsically more luminious, as selection effects may mean
that we don't see extreme [3.6]-[4.5] colours for distant sources in
the {\em GLIMPSE} data.

Another possible reason for the unexpectedly low detection rate for
new 6.7-GHz methanol masers could be that many of the {\em GLIMPSE}
sources in category A and B have previously been observed in either
targeted or untargeted searches.  A number of regions of the Galactic
plane have previously been subjected to untargeted searches
\citep{C96,EVM+96,SKHKP02}.  Approximately one quarter of the targeted
{\em GLIMPSE} sources were within these regions.  It could be argued
that these should have been excluded as part of the selection process,
however, they were included to see if they would lead to the discovery
of any new sources in the regions.  Since interstellar masers are
variable it is likely that a single epoch untargeted search will miss
some sources which happen to be beneath the detection threshold at the
time of the search.  The fact that no new masers were found in the
regions that had previously been subject to untargeted searches
suggests that the number of 6.7-GHz methanol masers missed in
untargeted searches due to variability is low.  The main targeted
searches have been toward previously known main-line OH maser sources
and {\em IRAS} sources meeting the \citet{WC89} ultra-compact \ionhy\/
region criteria.  The detection rate of 6.7-GHz methanol maser sites
towards previously known main-line OH masers resulted in very high
detection rates \citep[e.g.][]{M91,CVEWN95}, so we need only consider
the {\em IRAS}-based searches here.  Of the 200 targeted {\em GLIMPSE}
sources, 77 are within 30 arcseconds of an {\em IRAS} source.
However, only six of these {\em IRAS} sources meet the
\citeauthor{WC89} criteria of Log$_{10}(S_{60}/S_{12}) \ge 1.30$ \&
Log$_{10}(S_{25}/S_{12}) \ge 0.57$, and of these six none are
associated with detected 6.7-GHz methanol masers.  This means that few
of the {\em GLIMPSE} source positions targeted here would have been
targeted in previous {\em IRAS}-based searches.

The investigations above demonstrate that the unexpectedly low
detection rate of new 6.7-GHz methanol masers is not due to flaws in
the selection process for the target sources.  The most likely other
reasons for the low detection rate are either that the target sources
are primarily not regions of high-mass star formation, or that they
are regions of high-mass star formation prior to or post the 6.7-GHz
methanol maser evolutionary phase.  That only six of the 200 target
{\em GLIMPSE} sources are within 30~arcseconds of an {\em IRAS} source
with ultra-compact \ionhy\/ region colours would appear to rule out
the possibility that they are post the 6.7-GHz methanol maser
evolutionary phase.  While the possibility that they are not high-mass
star formation regions cannot be ruled out, their extreme mid-infrared
colours suggest that many may be very young high-mass star formation
regions, prior to the evolutionary phase associated with 6.7-GHz
methanol masers.

From the large number of previously known sources detected, it is
clear that that the estimates of E06 are conservative and that masers
separated by more than 3.5~arcmin from the {\em GLIMPSE} source are
frequently detected.  This is particularly true for the stronger
6.7-GHz methanol masers which in some cases have been detected when
separated by more than 30~arcmin from the {\em GLIMPSE} source.  The
results presented in Tables~\ref{tab:det} \& \ref{tab:maserprop} show
that sources with peak flux density less than 10~Jy are detected up to
a maximum separation of about 5~arcmin, increasing to about 8~arcmin
for a peak flux density less than 100~Jy.  The current search excluded
from the target list all {\em GLIMPSE} sources within 3.5 arcmin (half
the FWHM of the Mt Pleasant antenna beam at 6.7~GHz) of known methanol
maser sources.  Given that many methanol maser sources are moderately
strong this exclusion radius should probably be increased by 50 or 100
per cent in future similar searches to reduce the number of
redetections of known sources.

\subsection{Long-term variability}

Table~\ref{tab:longtermvar} compares the number of spectral features
identified in pre-1995 spectra of 6.7-GHz methanol masers taken from
the literature (where available) to those in
Fig.~\ref{fig:detections}.  Determining which spectral features have
appeared or disappeared is quite subjective.  For example, for sources
where there are a large number of spectral features in a relatively
small velocity range (e.g. G339.88-1.26 or G23.01-0.41) it is
impossible to assess from single dish spectra whether weaker spectral
features have appeared or disappeared.  In other cases the
observations in this paper have a poorer signal to noise ratio than
those from \cite{CVEWN95} and determining whether a spectral feature
has disappeared, or is simply too weak to identify in the 2005
spectrum is not possible.  Despite these limitations it is possible to
use the available spectra to make a crude estimate of the number of
spectral features which have appeared and disappeared.  The pre-1995
spectra for the 22 sources listed in Table~\ref{tab:longtermvar}
contain 156 identifiable spectral features.  Of the 22 sources, 10
showed no change in the number and velocity of spectral features in
the 2005 spectra.  For the other 12 sources there were either one or
two spectral features identified as having either appeared or
disappeared.  We can obtain a crude estimate of the average lifetime
of an individual 6.7-GHz methanol maser spectral feature from the
percentage of spectral features that either appear or disappear over
the given period.  The results summarised in
Table~\ref{tab:longtermvar} suggest that around 6 per cent (taking the
mean of the appear and disappear numbers) of 6.7-GHz methanol maser
features in a spectrum will change over a ten year period, implying an
average lifetime of approximately 150 years.  This is very much longer
than the typical lifetime of water masers, implying that the
environment in which methanol masers arise must be much more stable.
It also means that it should be possible to track individual methanol
maser features in VLBI proper motion experiments on timescales of
decades.

\begin{table*}
  \caption{Comparison of the number of 6.7-GHz methanol maser spectral 
    features observed in 2005 to observations made more than a decade ago.
    References : a = \citet{CVEWN95}; b = \citet{CVE95}; c = \citet{EVM+96};
    d = \citet{E96}.} 
  \begin{tabular}{lcccc} \hline
    {\bf Methanol} & \multicolumn{3}{c}{\bf Number of}         &  {\bf Reference} \\ 
    {\bf maser}    & \multicolumn{3}{c}{\bf Spectral Features} & \\
    {\bf name}     & {\bf pre-1995} & {\bf appeared 2005} & {\bf disappeared 2005} & \\ [2mm] \hline
    G\,$305.20\!+\!0.21$ & 3  & 1 & 1 & a \\ 
    G\,$305.21\!+\!0.21$ & 7  & 0 & 0 & a \\
    G\,$305.25\!+\!0.25$ & 2  & 0 & 0 & a \\
    G\,$316.81\!-\!0.06$ & 7  & 1 & 2 & b \\ 
    G\,$320.23\!-\!0.29$ & 11 & 0 & 1 & b \\ 
    G\,$323.46\!-\!0.08$ & 2  & 0 & 1 & a \\ 
    G\,$323.74\!-\!0.26$ & 15 & 1 & 1 & a \\ 
    G\,$326.66\!+\!0.52$ & 2  & 2 & 0 & c \\ 

    G\,$328.24\!-\!0.55$ & 10 & 0 & 2 & a \\ 
    G\,$328.25\!-\!0.53$ & 9  & 0 & 0 & a \\
    G\,$331.56\!-\!0.12$ & 7  & 0 & 0 & a \\
    G\,$337.15\!-\!0.40$ & 3  & 0 & 0 & d  \\
    G\,$337.39\!-\!0.20$ & 6  & 2 & 0 & d  \\
    G\,$339.88\!-\!1.26$ & 15 & 0 & 2 & a \\ 
    G\,$23.01\!-\!0.41$  & 14 & 0 & 1 & a \\ 
    G\,$25.82\!-\!0.17$  & 7  & 0 & 0 & d  \\
    G\,$29.86\!-\!0.04$  & 7  & 0 & 0 & a \\
    G\,$30.20\!-\!0.17$  & 6  & 0 & 1 & a \\ 
    G\,$30.70\!-\!0.07$  & 4  & 0 & 0 & a \\
    G\,$30.79\!+\!0.20$  & 8  & 0 & 0 & a \\
    G\,$30.82\!-\!0.05$  & 11 & 0 & 0 & a \\ \hline
  \end{tabular}
  \label{tab:longtermvar}
\end{table*}


\section{Conclusions} \label{sec:concl}

The {\em GLIMPSE} point source catalogue has been used to identify
regions likely associated with high-mass star formation at the
evolutionary phase when 6.7-GHz methanol masers occur.  A search of
the 200 most extreme sources meeting the selection criteria of E06
detected only nine new 6.7-GHz methanol masers, significantly less
than expected.  This implies that the most extreme sources meeting the
E06 criteria are less likely than average to have an associated
methanol maser, they may be associated with high-mass star formation
prior to the 6.7-GHz methanol maser evolutionary phase.

Comparison of the spectra of the previously known 6.7-GHz methanol
masers with those observed more than a decade ago (where available)
shows that over a ten year period about 6 per cent of the spectral
features will disappear and a similar number will appear.  This
implies the average lifetime of a methanol maser feature is
approximately 150 years, much longer than typically observed for
22-GHz water masers.

\section*{Acknowledgements}

Financial support for this work was provided by the Australian
Research Council.  This research has made use of NASA's Astrophysics
Data System Abstract Service.  This research has made use of data
products from the \glim\ survey, which is a legacy science program of
the {\em Spitzer Space Telescope}, funded by the National Aeronautics
and Space Administration.

\end{document}